\documentclass[aps,pra,twocolumn,preprintnumbers]{revtex4}

\usepackage{graphicx}  
\usepackage{subfigure}
\usepackage{multirow}
\usepackage{float}
\usepackage{fancyhdr}
\usepackage{longtable}
\usepackage{parskip}
\usepackage[T1]{fontenc}
\usepackage{dcolumn}   
\usepackage{amsmath}
\usepackage[dvipsnames]{xcolor}
\usepackage{soul}
\usepackage[utf8]{inputenc}
\usepackage[english]{babel}
\usepackage{bm}        
\usepackage{amsfonts}  
\usepackage{amsmath}   
\usepackage{amssymb}   

\newcommand{\exv}[1]{\left\langle #1 \right\rangle}
\newcommand{\pwisein}{\left\{ \begin{array}{ll}}
\newcommand{\pwiseout}{\end{array}\right.}

\begin{document}

\title{The effect of chaos on the simulation of quantum critical phenomena in analog quantum simulators}


\begin{abstract}
We study how chaos, introduced by a weak perturbation, affects the reliability of the output of analog quantum simulation.  As a toy model, we consider the Lipkin-Meshkov-Glick (LMG) model.  Inspired by the semiclassical behavior of the order parameter in the thermodynamic limit, we propose a protocol to measure the quantum phase transition in the ground state and the dynamical quantum phase transition associated with quench dynamics. We show that the presence of a small time-dependent perturbation can render the dynamics of the system chaotic. We then show that the estimates of the critical points of these quantum phase transitions, obtained from the quantum simulation of its dynamics, are robust to the presence of this chaotic perturbation, while other aspects of the system, such as the mean magnetization are fragile, and therefore cannot be reliably extracted from this simulator. This can be understood in terms of the simulated quantities that depend on the global structure of phase space vs. those that depend on local trajectories.
\end{abstract}

\author{Karthik Chinni$^{1}$, Pablo M. Poggi$^{1}$, Ivan H. Deutsch$^{1}$}

\affiliation {\it $^{1}$Center for Quantum Information and Control (CQuIC), Department of Physics and Astronomy,
University of New Mexico, Albuquerque, New Mexico 87131, USA}

\date{March 3, 2021}

\maketitle 
\section{Introduction}
Quantum simulators promise solutions to a wide variety of problems in many-body system physics that cannot be solved efficiently by analytic approximation or classical numerical simulation~\cite{Cirac, Nori, Hensgens, Aspuru, Garcia}. Quantum simulation is widely seen as a potential application for  noisy intermediate-scale quantum (NISQ) devices, comprised of $\sim 100$ qubits, but unable to perform fault-tolerant quantum computation~\cite{Preskill}. The absence of error correction in these NISQ-era quantum simulators raises an important question about their reliability. That is, can one trust the output of these devices, which are subjected to noise and imperfections~\cite{Deutsch2012, Deutsch2020, Sarovar}? Errors  arise from a variety of causes, which are typically  characterized as control errors due to miscalibration or inhomogeneities, uncontrolled classical noise, and decoherence due to entanglement with a quantum reservoir. Recently there have been various studies about the reliability and quantum advantage of NISQ devices in the presence of such errors for applications including simulation~\cite{ Pablo, Monroe2017, Lukin2017, Bloch2012, Safavi2018}, optimizations~\cite{Perruzo2014}, and random sampling~\cite{Arute2019}.  

Another source of error is ``dynamical complexity.'' NISQ devices, even in the circuit model, are fundamentally analog devices, operating with a continuous set of quantum maps~\cite{Deutsch2020}. As such, dynamical instabilities such as bifurcations and the onset of chaos can translate into a quantum map that leads to a proliferation of errors.  Indeed, quantum chaos can be characterized as the hypersensitivity of quantum dynamics to external perturbations in the Hamiltonian~\cite{Peres, Schack} and thus we expect that quantum chaos may limit our ability to reliably extract the desired output of analog (not digitally error corrected) quantum simulators, especially in cases where such devices are expected to solve hard problems.

The effect of quantum chaos on quantum computing was considered in the early 2000s~\cite{Georgeot, Song, Flambaum, Braun}. Georgeot \textit{et al}. studied how static imperfections such as qubit level fluctuations and residual (always on) interactions between qubits can lead to quantum chaos that destroys the register states of the system in the absence of error correction~\cite{Georgeot}. Song \textit{et al}. studied the effects of errors in quantum algorithms while simulating the chaotic regime of the kicked-rotor model and showed that the error in the diffusive constant, which is associated with chaotic dynamics, grows exponentially with the system size~\cite{Song}. More recently, dynamical complexity in quantum simulation has come to the fore. In a series of papers, Heyl {\em et al.}~\cite{Heyl2019} and Sieberer \textit{et al.}~\cite{Sieberer2019} studied the simulation of a Ising-type Hamiltonian through Trotterization in a gate-model.  In this approximation, the unitary map consists of a series of Floquet maps describing the dynamics of a delta-kicked system, which is quantum-chaotic in a particular parameter regime. They have shown that the resulting magnetization errors increase sharply in this quantum-chaotic regime. 

In this work we study the effect of quantum chaos on different aspects of quantum phase transitions (QPTs) in a quantum simulation. For models describable in the thermodynamic limit by mean-field theories~\cite{Bapst2012}, QPTs are often associated with bifurcations in the phase space dynamics that governs the order parameters.  These bifurcations lead to unstable fixed points and separatrix lines, and we expect chaos to develop in their vicinity in the presence of small perturbations.  Thus, it is natural to consider how quantum chaos affects the reliability of the quantum simulation in such models.  We study this in the simplest paradigm, the Lipkin-Meshkov-Glick (LMG) model, which describes the anharmonic motion of a collective spin.

The LMG model undergoes a ground state quantum phase transition (GSQPT) in the mean magnetization ~\cite{Santos2016}, and a dynamical quantum phase transition (DQPT) in the time-averaged magnetization of the quench dynamics~\cite{Zunkovic2018}.  We will characterize the emergence of chaotic dynamics in the LMG model when a weak time-dependent perturbation is applied and analyze the resulting robustness of the simulation protocols. Importantly, we will show that a key quantity such as the time-averaged magnetization, extracted from this quantum simulation, can be  highly sensitive to chaos in certain regimes. Nonetheless, we find that other aspects of the quantum simulation, such as the estimation of critical points for both the GSQPT and the DQPT, are robust to the presence of the perturbation and chaos. This difference in fragility vs. robustness can be explained by the difference between ``fine-grained'' and ``coarse-grained'' information being extracted from the quantum simulator.

The remainder of this manuscript is organized as follows. In Sec. II we review the LMG model, and analyze its thermodynamic limit to explain the underlying mechanism associated with both the GSQPT and the DQPT present in this model. Following this, in Sec. III we present a protocol that allows us to extract the critical points associated with both the GSQPT and the DQPT from a unitary quantum simulator. Then, in Sec. IV.A we characterize the chaotic dynamics emerging from the simulation of the LMG model due to the presence of a background time-dependent perturbation. Finally, in Sec. IV.B we analyze the robustness of two particular quantities, the time-averaged magnetization and the critical point estimates of the GSQPT and the DQPT, to the presence of a time-dependent perturbation. 

\begin{figure*}[t!]
\centering
\includegraphics[width=0.95\textwidth]{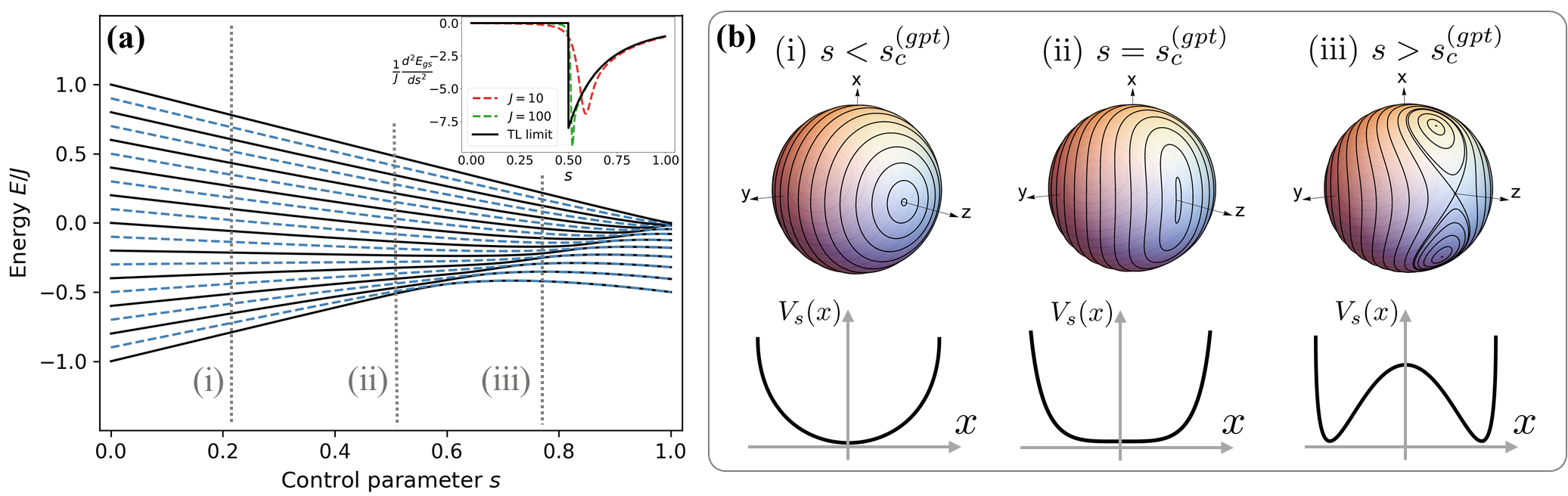}
\caption{\label{setup} 
(a) Energy spectrum of the LMG Hamiltonian, Eq. (\ref{ec:hami_lmg}), as a function of the control parameter $s$ for $N=20$ ($J=10$). Thick and dashed lines indicate positive and negative parity eigenstates respectively. Inset shows the second derivative with respect to $s$ of the lowest energy eigenvalue, which exhibits a discontinuity at the critical point of the GSQPT, $s=s_c^{(gpt)}=0.5$ in the thermodynamic limit. (b) Classical phase-space portraits of the mean-field dynamics of the LMG model at various values of the control parameter $s$ (i)-(iii) corresponding to the dashed lines in (a). For each case, we display the effective one-dimensional potential $V_s(x)$, which changes from a single well for $s<s_c^{(gpt)}$ to a double well for $s>s_c^{(gpt)}$.   }
\label{fig:fig1}
\end{figure*}

\section{Quantum Phase Transitions in the LMG Model} 
The LMG model describes an infinite-range transverse field Ising model on a completely connected graph. We consider the specific case of the Curie Model~\cite{Kochmanski2013} governed by the Hamiltonian
\begin{equation}
    H=-\frac{h}{2}\sum_{i=1}^{N} \sigma_{z}^{(i)}-\frac{\gamma}{4N}\sum_{i,j=1}^{N} \sigma_{x}^{(i)} \sigma_{x}^{(j)} =-hJ_{z}-\frac{\gamma}{N}J_{x}^{2}
\label{ec:intro_hami_lmg}
\end{equation}
where $J_{\alpha}=\sum_{i=1}^{N}\frac{\sigma_{\alpha}^{(i)}}{2}$ are the collective angular momentum operators. While originally studied in the context of nuclear physics~\cite{Lipkin1965}, the LMG model has applications in a wide variety of contexts, including the dynamics of two-mode Bose Einstein condensates for applications of quantum metrology~\cite{Julia2012, Strobel2014,  Pezze2018}, studies of quantum chaos~\cite{Tomkovic2015, Mourik2018}, and quantum simulation~\cite{Monroe2017, Lynse2020,Xu2020}. For the remainder of this manuscript, we scale the energy and parametrize the LMG Hamiltonian with a single parameter $s$ according to
\begin{equation}
H(s)=-(1-s)J_{z}-\frac{s}{N}J_{x}^{2},
\label{ec:hami_lmg}
\end{equation} 
\noindent with $0\leq s\leq1$. The total angular momentum $J^{2}$ is conserved, allowing us to focus on the dynamics within the symmetric subspace with $J=N/2$, which is spanned by the $2J +1= N+1$ Dicke states. 

The LMG model has a second-order continuous GSQPT at $s=\frac{1}{2}$ in the thermodynamic limit, $N\rightarrow \infty$, resulting from the competition between the external magnetic field inducing paramagnetic order and the spin-spin interactions inducing ferromagnetic order~\cite{Santos2016}. A signature of this phase transition can be seen on the energy spectrum, shown in Fig. \ref{fig:fig1}a, from the closing of the energy gap between the ground state and the first excited state at around $s=s_{c}^{(gpt)}(N)$, which is different from the thermodynamic-limit value $s_{c}^{(gpt)}=\frac{1}{2}$ due to finite-size effects. In this phase transition, the ground state changes its character continuously from paramagnetic to ferromagnetic in nature. The order parameter associated with this phase transition is the magnetization along the $x$-axis, $\langle\psi_{gs}|J_{x}|\psi_{gs}\rangle$. This order parameter is zero in the paramagnetic phase and nonzero in the ferromagnetic phase. The presence of this second order GSQPT can also be seen by noting the nonanalyticity in the second-order derivative of the ground-state energy at the critical point as shown in the inset of Fig. \ref{fig:fig1}a. In addition, the LMG model has a continuum of excited-state quantum phase transitions (ESQPTs) along the critical line, $E_{\text{ESQPT}}/J=-(1-s)$, in thermodynamic limit \cite{Caprio2008, Santos2016}. A signature of these ESQPTs can be seen on the spectrum in the local divergence of the density of states along this line.

The LMG model also exhibits a DQPT in a order parameter associated with nonequilibrium dynamics \footnote{In general, there are two different kinds of DQPTs. The first type is
defined by the appearance of the singularities in the time
evolution of the survival probability or Loschmidt echo,
referred to as DQPT-LE. The second type is defined
by the presence an order parameter that is zero in one
phase and nonzero in the other phase, referred to as
DQPT-OP. In this work, we focus on
quantum simulation of the DQPT-OP (which we refer
to as just DQPT). For more details, refer \cite{Zunkovic2018}.}. In particular, the order parameter associated with this phase transition is the time-averaged magnetization along the $x$-axis,
\begin{equation} 
\overline{\langle J_{x}\rangle}=\lim\limits_{T\rightarrow\infty}\frac{1}{T}\int_{0}^{T}dt\;\langle J_{x}\rangle(t).
\label{DQPT-OP}
\end{equation}
 In the paradigm of DQPT, an initial state $|\psi_{0}\rangle$, which is the ground state of the Hamiltonian $H(s_{i})$ for an initial value of the control parameter is time-evolved under a quenched Hamiltonian $H(s_{f})$, and the long-time dynamics of the state is analyzed. In this work, we fix the initial value of the control parameter to $s_{i}=1$, and choose the initial state to be one of the fully polarized states along the $x$-axis. At  $s=s_{c}^{(dpt)}=\frac{2}{3}$, the dynamical behavior of the state changes from precessesing around the external magnetic field to evolving closely around its initial position under the action of the quenched Hamiltonian, resulting in a DQPT ~\cite{Zunkovic2018, Monroe2017}. The order parameter, Eq. (\ref{DQPT-OP}), is zero for $s_{f}<s_{c}^{(dpt)}$ and has a nonzero value for $s_{f}>s_{c}^{(dpt)}$. 

\subsection{Semiclassical description of phase transitions} \label{sec:semiclassical}

The LMG model is an example of a quantum mean-field model in which the thermodynamic limit, $N\rightarrow \infty$, is equivalent to the dynamics of the mean field~\cite{Bapst2012}.  We obtain these dynamics from the expectation values of the Heisenberg equations of motion, neglecting all fluctuations by approximating $\langle A B \rangle \approx \langle A  \rangle \langle B \rangle $.  The resulting mean-field dynamics are thus equivalent to the classical dynamics of a massless ``top.'' Defining the unit vector $(X,\:Y,\:Z)\equiv \lim_{J\rightarrow \infty} \frac{1}{J}(\exv{J_{x}},\:\exv{J_{y}},\:\exv{J_{z}})$, the classical equations of motion are~\cite{Milburn1997}
\begin{align}
\begin{split}
\frac{dX}{dt}	&=(1-s)Y , \\
\frac{dY}{dt}	&=-(1-s)X+sXZ ,\\
\frac{dZ}{dt}	&=-sXY .\\
\end{split}
\label{ec:lmg_class}
\end{align}
The above equations of motion can be solved analytically. As expected, $\frac{d}{dt}(X^{2}+Y^{2}+Z^{2})=0$, indicating that the classical LMG Hamiltonian describes a system with a single degree of freedom, whose phase space is the unit sphere. Since the LMG Hamiltonian is time-independent, the energy of the system is conserved, and thus the system is integrable and motion is regular.
 
 The essential features of the classical (mean-field/thermodynamic) limit of the LMG Hamiltonian can be captured using the phase-space diagrams, such as the ones shown for different values of $s$ in Fig. \ref{fig:fig1}b.
 Note that for values of $s$ smaller than $\frac{1}{2}$, all the phase-space trajectories precess around the $z$ axis. At $s = \frac{1}{2}$ a bifurcation occurs. The topology of the phase-space trajectories changes as the fixed point located at $\theta=0$, which is stable for $s<\frac{1}{2}$, becomes unstable for $s>\frac{1}{2}$ and two new stable fixed points are formed, which are located at $(\theta,\phi)=\bigl(\pm \cos^{-1}(\frac{1-s}{s}),0\bigr)$ (Here $\theta$ and $\phi$ are the usual angular coordinates on the sphere, $\theta = \cos^{-1}(Z)$ and $\phi = \tan^{-1}( Y/X)$). This major reconfiguration of phase-space structure signifies the onset of the GSQPT. As a result of bifurcation of the fixed point, there are two kinds of trajectories present on the phase space for $s>\frac{1}{2}$, the ones that precess around the stable fixed points, and the ones that revolve around the whole sphere. These two kinds of trajectories are separated by a separatrix layer, which includes the unstable fixed point located at $\theta=0$. As the value of $s$ is increased from $\frac{1}{2}$ to $1$, these stable fixed points move farther away from the unstable fixed point and the separatrix layer bounds more trajectories on the phase space.
 
\begin{figure}[t]
\centering
\includegraphics[width=\linewidth]{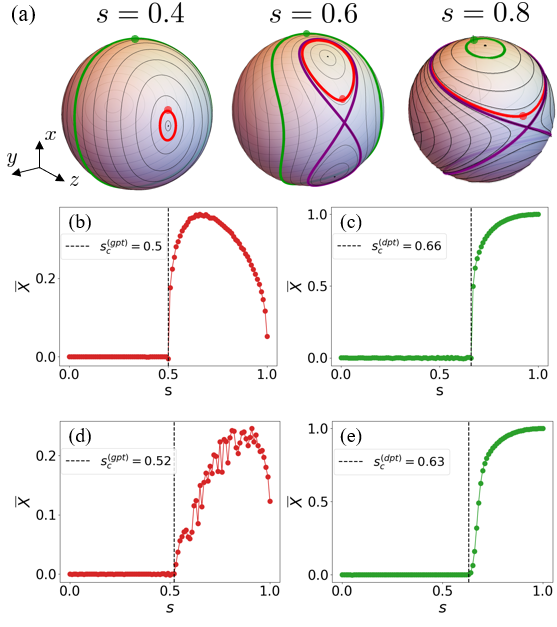}
\caption{(a) Classical phase-space portraits of the LMG Hamiltonian for different values of $s$. Red and green curves indicate the evolution corresponding to the GSQPT and DQPT bifurcation protocols, respectively (initial conditions are marked as dots). The separatrix trajectory (purple) separates rotations from librations on the phase space. (b)-(c) Classical bifurcation diagram showing the long-time average of $X=J_x/J$ for both the GSQPT (b) and DQPT. (c) Protocols as a function of the control parameter $s$. A sharp change is seen at the corresponding critical points $s_c^{(gpt)}=0.5$ and $s_c^{(dpt)}=2/3$. (d)-(e) Same as (b)-(c) but for quantum evolution with $N=200$ ($J=100$). Here, the initial conditions are spin coherent states centered roughly at the points of the classical protocols (see text for details).}
\label{fig:fig2}
\end{figure}

The DQPT also admits a clear description in terms of the classical trajectories. 
The system is first initialized in the fully polarized state along the $x$-axis, which is a ground state at $s=1$, and then allowed to evolve under the action of the quenched Hamiltonian, $H(s_{f})$. In the thermodynamic limit, as the value of $s_{f}$ is varied, the position of the initial state changes with respect to the separatrix layer, resulting in different dynamical phases \cite{Zunkovic2018}. For $s_{f}<\frac{2}{3}$, the state precesses around the $z$-axis resulting in a zero value for the order parameter (see Fig. \ref{fig:fig2}a and Fig. \ref{fig:fig2}b). At the dynamical critical point, $s_{f}=\frac{2}{3}$, the state is exactly located on the separatrix layer, while for $s_{f}>\frac{2}{3}$, the state precesses around one of the stable fixed points, as shown in Fig. \ref{fig:fig2}c, resulting in a nonzero value of the order parameter for these $s$ values.

We gain further insight into the dynamics in the semiclassical limit by mapping the LMG to a problem of a fictitious particle moving in a one-dimensional well in the limit $N\gg1$~\cite{BEC, Julia2012, joseph2_well, Santos2016, heiss_well, hp_well, reibero_well}. Following \cite{Julia2012}, one can derive a time-dependent Schrodinger equation in the Fock basis, obtained by expressing the spin operators in the Schwinger representation. Ignoring terms of order $\mathcal{O}(N^{-3})$ and smaller,  the time-dependent Schrodinger equation along the $x$-axis can be written as 
\begin{align}
\begin{split}
i\frac{1}{N}\frac{\partial}{\partial t}\psi(x,t)&=-\frac{1}{N^2}(1-s)(\partial_{x}\sqrt{1-x^{2}}\partial_{x})\psi(x,t) \\
& -\frac{1}{4}\biggl(2(1-s)\sqrt{1-x^{2}}+sx^{2}+(1-s) \\
& \biggl(\frac{2}{N\sqrt{1-x^{2}}}-\frac{1}{N^2}\frac{1+x^{2}}{(1-x^{2})^{3/2}}\biggr)\biggr)\psi(x,t)
\end{split}
\label{ec:semiclassical_potential}
\end{align}
This can be interpreted as the Schrodinger equation for a particle with a position-dependent mass, moving in a 1D potential $V_s(x)= -\frac{1}{2}(1-s)\sqrt{1-x^{2}}-\frac{1}{4}sx^{2}$, plus $N$-dependent corrections. Importantly, as shown in Fig. \ref{fig:fig1}b, this potential well $V_s(x)$ changes its shape from a single well to a double well at $s=\frac{1}{2}$ as $s$ is increased from $0$ to $1$ in this large-$N$ limit. 

Classical motion of a particle in this potential also provides intuition to explain the presence of the GSQPT and the DQPT in the LMG. At $s=1/2$ the potential changes its shape from a single well to a double well, corresponding to the critical point of the GSQPT in the thermodynamic limit, which is accompanied by the spontaneous symmetry breaking in accordance with Landau-Ginzberg theory \cite{Goldenfeld}.
Note that for a finite value of $N$, the critical point, $s^{(gpt)}_{c} (N)$, moves to a higher value of $s$ compared to the thermodynamic
limit value, $s^{(gpt)}_{c}$, due to 
the finite-size zero-point energy of the ground state, which
is higher than the barrier height of the shallow
well for finite-$N$ values. Therefore, one can identify the
value of $s$ at which the zero-point energy equals the barrier
height of the double well as the finite-size critical
point. For more details refer to Appendix \ref{sec:FSS}, where we derive the finite-size scaling of the ground-state critical point.

The DQPT can be understood as a transition across the separatrix line.  For the specially chosen initial condition, at $s=2/3$ the system in the classical description is exactly on the separatrix in phase space; its energy is equal to the height of the barrier of the double well.  For $s<2/3$ the pseudoparticle is above the barrier and executes librations in the double well. This corresponds to the top precessing around the external magnetic field in the LMG model, resulting in the order parameter $\overline{X}=0$ in the long-time limit. In the second dynamical phase, for $s>\frac{2}{3}$, the energy is lower than the barrier height of the double-well potential. As a result, the pseudoparticle becomes trapped on one side of the double well, corresponding to precession in phase space around one of the stable fixed points, and $\overline{X} \neq 0$.  This explains why the critical value of $s$ is greater for the DQPT than the GSQPT. The double-well forms first  at the bifurcation point, yielding the GSQPT; for larger values of $s$ the barrier height becomes equal to the energy of the specially chosen initial condition yielding the DQPT.

Moreover, the double-well potential in the classical picture also explains the presence of the ESQPTs \cite{Santos2016}. A classical particle with the energy equal to the separatrix energy spends a very long time in the vicinity of the unstable fixed point, which implies that the probability of finding the associated quantum state is very high in such neighborhood. Hence, the state undergoing an ESQPT is localized around the unstable fixed point. Since the barrier height of the double well increases with the value of the control parameter, the energy of the state undergoing ESQPT increases as well, and the energy of the phase space separatrix is $E_{\text{ESQPT}}/J=-(1-s)$, in the classical limit.
\section{Accessing QPTs with dynamical quantum simulators}
We are interested in studying QPTs in a quantum simulator whose ideal operation involves implementing a desired unitary transformation on a chosen initial state, followed by measurement and data processing to extract a desired output~\cite{Monroe2017, Lynse2020}. On this type of device, some states, such as spin-coherent states, can be prepared with high fidelity, and the expectation values of certain observables are accessible from measurements after the system evolves under the Hamiltonian of interest for a chosen duration. With this setting in mind, we seek to study the critical phenomena associated with both the GSQPT and the DQPT in the LMG model, and in particular we want to extract the critical points associated with these phase transitions. Recently, protocols have been proposed to identify the critical points in some of the Ising-like models by measuring out-of-time-ordered (OTO) correlators \cite{OTOCs}, or studying the spectrum of multiple quantum coherences \cite{MQC}, or through adiabatic quenches \cite{quench}. In this section, we propose a protocol for identifying the critical point of the GSQPT, which only requires us to have the ability to prepare a particular spin-coherent state and the ability to measure the collective magnetization in the longitudinal direction, $\langle\psi(t)|J_{x}|\psi(t)\rangle$, as a function of time. We first motivate our protocol based on the classical phase-space trajectories and then provide justification on why this protocol would still work in the case of a finite-dimensional system. Moreover, we will see that changing the initial state in the above mentioned protocol also allows us to measure the order parameter associated with the DQPT and thereby allowing us to extract the DQPT critical point.

\subsection{Classical bifurcation}
As discussed in Sec. \ref{sec:semiclassical}, in the thermodynamic limit a stable fixed point at $\theta=0$ bifurcates into two other stable fixed points resulting in the change of topology of the phase-space trajectories at the critical point of the GSQPT, in accordance with the Landau-Ginzberg theory. Therefore, as shown in Fig. \ref{fig:fig1}b, the single-well potential changes into a double-well potential at $s=\frac{1}{2}$ as $s$ is increased from $0$ to $1$. Based on this semiclassical picture, we propose a protocol to identify the critical point of the GSQPT by computing the time-averaged magnetization, $\overline{X}$, of an initial condition located at an angle slightly off the $z$-axis, $(\theta_{0}=\epsilon_{TL},\phi_{0}=0)$, where $\epsilon_{TL}$ is a small angle. We show in the Appendix \ref{sec:classical bifu} that the critical point estimate of the GSQPT is robust to perturbations in this angle. In this work, we choose $\epsilon_{TL}=\frac{\pi}{60}$. This time-averaged magnetization, when plotted as a function of $s$, exhibits a pitchfork bifurcation at the GSQPT critical point \cite{bifu1}, as shown in Fig. \ref{fig:fig2}b. In this figure, the zero values of the time-averaged magnetization correspond to the values of $s$ where the initial condition is oscillating in the single-well potential centered at $\theta= 0$ as shown in Fig. \ref{fig:fig1}b. The nonzero values correspond to the values of $s$ where the initial condition's dynamics is constrained to one side of the double well centered at $(\theta =\cos^{-1}(\frac{1-s}{s}),\; \phi=0 \bigr)$. 

Likewise, the time-averaged magnetization for the initial condition located at $(\theta_{0},\phi_{0})=(\frac{\pi}{2},0)$, which is the order parameter associated with the DQPT, also shows a pitchfork bifurcation at $s=\frac{2}{3}$, as shown in Fig. 2c, because the initial state can access the whole well, single-well or double-well potential, for $s<\frac{2}{3}$ while its dynamics is constrained to one side of the double well for $s>\frac{2}{3}$ \cite{Zunkovic2018}. Note that this bifurcation is expected because $\overline{X}$ is the order parameter for the DQPT, which is zero in one phase and nonzero in the other phase by definition. As a result, this bifurcation protocol can be used to determine the critical points of both the GSQPT and the DQPT in the thermodynamic limit provided appropriate initial conditions are used, $(\theta_{0}=\epsilon_{TL},\phi_{0}=0)$ for GSQPT and $(\theta_{0}=\frac{\pi}{2},\phi_{0}=0)$ for DQPT. Note that, as a consequence of the parity symmetry, one could also construct the negative branch associated with both the bifurcation diagrams in Figs. \ref{fig:fig2}b and c by starting the protocol in initial states that are rotated by $\pi$ about the $z$-axis from the initial states corresponding to the positive branch. 

Since LMG model is a system with one degree of freedom, energy conservation allows us to gain complete knowledge of the system. In particular, we can derive an analytic expression for the time-averaged magnetization for all the initial conditions of the form $(\theta_{0},\phi_{0}=0)$, which is given by
\begin{align}
\overline{X}&= \begin{cases} 
      0 & s < \frac{1}{1+\cos^{2}\bigl(\frac{\theta_{0}}{2}\bigr)} \\
      \frac{\pi}{2} \frac{\sin{\theta_{0}}}{K\bigl(\Lambda(\theta_{0},s)\bigr)} & s\geq \frac{1}{1+\cos^{2}\bigl(\frac{\theta_{0}}{2}\bigr)}
   \end{cases}
\end{align}
where $\Lambda(\theta_{0},s)=-\frac{4(1-s)}{s\sin^{2}(\theta_{0})}\bigl(\cos(\theta_{0})-\frac{1-s}{s}\bigr)$, and $K(x)$ denotes the complete elliptic integral of the first kind. We present the full derivation and further details in Appendix \ref{sec:classical bifu}. The time-averaged magnetization in the LMG model has been explored in various previous works \cite{Monroe2017,magnetization_formula2,magnetization_formula3} and a similar calculation is presented in Ref. \cite{magnetization_formula1}. Note that the above result also provides us with the bifurcation points along with their respective bifurcation curves for all initial conditions of the form $(\theta_{0},\phi_{0}=0)$. The two cases of interest for us are the $\theta_{0} =\epsilon_{TL}$ (GSQPT) and $\theta_{0}=\frac{\pi}{2}$ (DQPT), whose bifurcation points are given by  $(1+\cos^{2}(\frac{\epsilon_{TL}}{2}))^{-1} \simeq 0.5$ and $(1+\cos^{2}(\frac{\pi}{4}))^{-1}=\frac{2}{3}$ respectively. Furthermore, the above expression for the bifurcation point also explains its robustness with respect to the choice of initial $\theta$ for the GSQPT case, which can be inferred from the absence of the first order term in the Taylor expansion of $(1+\cos^{2}(\frac{\theta}{2}))^{-1}$ around $\theta=0$.

\subsection{Quantum bifurcation}
In this subsection, we show that even for a finite-dimensional system, the protocol described above allows us to estimate the finite-size critical points of both the GSQPT and the DQPT. For the GSQPT, a spin-coherent state centered at an angle slightly off the classical unstable fixed point, $(\theta=\epsilon(N), \phi_{0}=0)$, labelled by $|\theta=\epsilon(N),\phi_{0}=0\rangle$ is time-evolved under the LMG Hamiltonian for different values of $s$. Here, we set $\epsilon(N)=\frac{\pi}{60}+\frac{1}{\sqrt{N}}$, which accommodates the variance of the spin-coherent state and approaches $\epsilon_{TL}$ in the thermodynamic limit. The resulting time-average from the time-evolution, $\overline{X}=\overline{\langle J_{x}\rangle}/J$, is then analyzed as a function of $s$, as shown in Fig. \ref{fig:fig2}d. The finite-size critical point can be estimated as the value of $s$ in which $\overline{X}$ changes from zero to a nonzero value. An important point here is the choice of the averaging time, denoted by $T$, which in this case should be larger compared to the time-scales associated with the periods of oscillation localized in a single-well, but also smaller compared to the time scales associated with the tunnelling time between the two sides of the double well. This can be inferred from the following argument. 

Consider the expectation value of $\langle J_{x} \rangle$ averaged over time, $\overline{\langle J_{x} \rangle}$, which can be written as
\begin{align} 
\overline{\langle J_{x}\rangle(t)}&=\frac{1}{T}\int_{0}^{T}\text{dt}\langle\psi(0)|e^{iHt}J_{x}e^{-iHt}|\psi(0)\rangle \label{ec:quantum_bifu1} \\
&=\sum_{n,m}c_{n}^{*}c_{m}\langle u_{n}|J_{x}|u_{m}\rangle\frac{1}{T}\int_{0}^{T}\text{dt}\: e^{i(E_{n}-E_{m})t} \label{ec:quantum_bifu2}
\end{align}
where the initial state $|\psi(0)\rangle$ in Eq. (\ref{ec:quantum_bifu1})  has been expressed in the eigenbasis of $H$, $|\psi(0)\rangle=\sum_{n} c_{n}|u_{n}\rangle$ resulting in Eq. (\ref{ec:quantum_bifu2}). Note that the above expression becomes $\overline{\langle J_{x}\rangle}=\sum_{n} |c_{n}|^{2}\langle u_{n}|J_{x}|u_{n}\rangle$ for the averaging time $T>> \frac{2\pi}{E_{n}-E_{m}}$ because the time-averaging integral in Eq. (\ref{ec:quantum_bifu2}) becomes a Kronecker delta function, $\delta_{nm}$, assuming that the energy states are non-degenerate. Since $J_{x}$ is odd under the action of the parity operator, $e^{i\pi J_{z}}$, the above expression finally becomes zero for $T>> \frac{2\pi}{E_{n}-E_{m}}$. In the LMG Hamiltonian, the gap between opposite parity energy eigenstates decreases as $s$ is increased from $0$ to $1$ and starts to close exponentially with the system size in the ferromagnetic phase \cite{CUTs}, $s>s_{c}^{(gpt)}(N)$, signalling the onset of the GSQPT. Therefore, we find that choosing the average time $T$ such that its much larger than the typical inverse energy gap for $s<s_c$, and using the same value of averaging time, $T$, for all $s$ values does a good job at reproducing the expected bifurcation.\break\break
For the case of the DQPT, the ground state at $s=1$, which is a spin-coherent state centered at $(\theta=\frac{\pi}{2}, \phi_{0}=0)$, is time-evolved under the action of the LMG Hamiltonian for different values of $s$ to obtain the the bifurcation diagram shown in Fig \ref{fig:fig2}e. Since the true order parameter of the DQPT is being measured in this case, we expect this protocol to give us an estimate of the finite-size critical point for all cases.

\section{Chaos and sensitivity to perturbations in a quantum simulation}
\subsection{Chaos in dynamical quantum simulation}
\begin{figure}[t] 
\includegraphics[width=\linewidth]{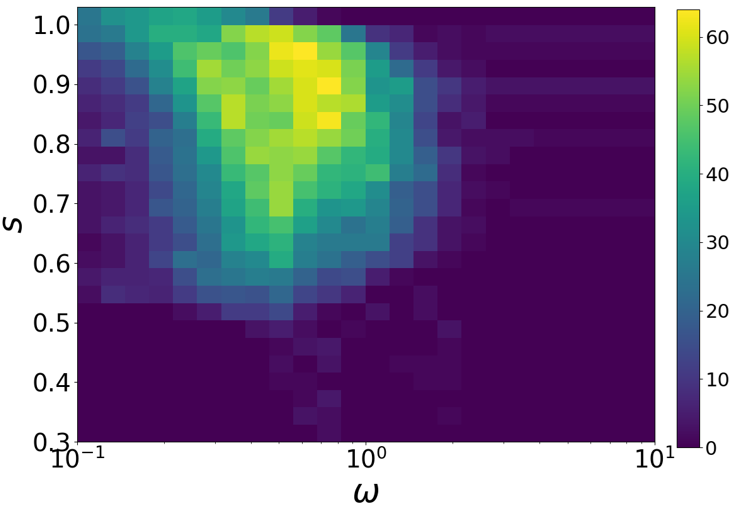}
\caption{Heat map showing the fraction of classical phase space of the LMG model that is chaotic for a range of frequencies, $\omega$, and control parameter, $s$. The perturbation amplitude is fixed at $\varepsilon_{0}=0.05$. The colorbar on the right relates different colors on the heat map with the percentage of phase space that is chaotic. See text for details.} 
\label{fig:fig3}
\end{figure}

The LMG Hamiltonian describes a system with one degree of freedom with no explicit time dependence implying that the classical system is integrable. This integrability can be broken upon addition of a perturbation that does not preserve the original symmetries of the system. For instance, in Ref. \cite{Lerose2018} the authors explore the impact of adding nearest-neighbor interactions in the LMG model. This perturbation, which breaks the permutational invariance of the system, leads to a new dynamical phase, where slight changes in the parameter of the Hamiltonian results in a very different dynamical behavior of the system. In this work we study the effect of a weak time-dependent perturbation that renders the system nonintegrable, with the goal of studying the impact of chaos on the quantum simulation of critical quantities associated with both the GSQPT and the DQPT. Here we choose a perturbation of a weakly-oscillating B-field along the $y$-axis, which results in the following effective Hamiltonian
\begin{equation}
H=-(1-s)\:J_{z}-\frac{s}{N}J_{x}^{2}-\varepsilon_{0}\cos(\omega t)J_{y}.\\
\label{ec: pert_hami}
\end{equation}
Mourik \textit{et al.} showed that a Hamiltonian of this form shows chaotic behavior for appropriate values of $\varepsilon_{0}$ and $s$ \cite{Mourik2018}. To evaluate the emergence of chaos in the perturbed system, we computed the fraction of phase space that becomes chaotic in the presence of this $J_{y}$ perturbation term for a range of frequencies and $s$ values, which is shown in Fig. \ref{fig:fig3}. Each data point on the heat map was obtained by identifying the fraction of initial conditions whose distance from their corresponding neighboring points separated exponentially for a given value of frequency and $s$ using the appropriate classical equations of motion. This data is in excellent agreement with the associated Poincar\'{e} sections, similar to that seen in Ref.~\cite{Mourik2018}. 

\begin{figure*}[t]
\centering
\includegraphics[width=0.9\textwidth]{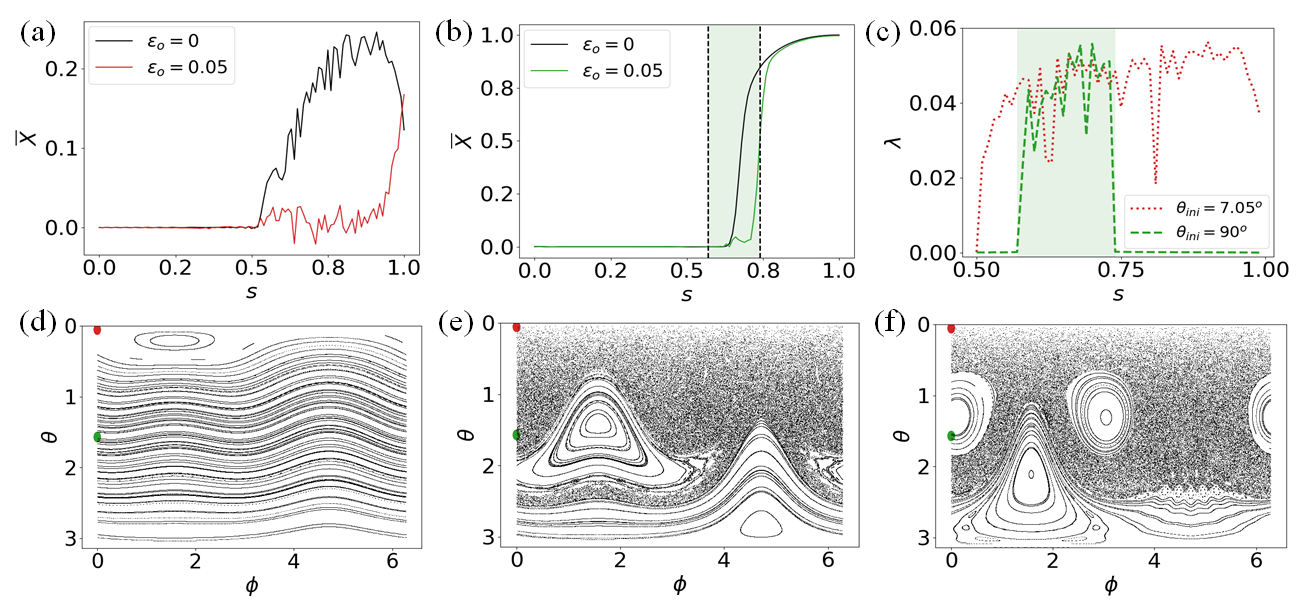}
\caption{The ideal and the perturbed bifurcation plots associated with the GSQPT (a) and the DQPT (b) are shown by the black and the red (a)/green (b) colored curves respectively as a function of the control parameter, $s$. It can be seen that time-averaged magnetization has been modified significantly for $s\gtrsim s_{c}^{(gpt)}(N)$ (a) / $ s_{c}^{(dpt)}(N) \lesssim s \lesssim 0.8$ (b), which corresponds to the values of $s$ where the associated initial condition is in the chaotic sea. The green shaded region highlights the values of the control parameter where the time-averaged magnetization for DQPT is fragile.  (c) Lyapunov exponents are plotted as a function of $s$ for initial conditions corresponding to GSQPT(dotted line) and DQPT(dashed line). The region where the DQPT initial condition has a positive Lyapunov exponent is highlighted by in green. Poincar\'{e} sections in the presence of chaotic perturbation are shown here for $s=0.3$ (d), $s=0.65$ (e) and $s=0.8$ (f) as a function of the coordinates of the classical phase-space, $\theta$ and $\phi$. The red and green circles label initial conditions associated with the GSQPT and  the DQPT respectively}
\label{fig:fig4}
\end{figure*}

Note that this system becomes chaotic mostly only for $s>1/2$ as shown in Fig. \ref{fig:fig3}. This can be explained by the presence of the separatrix layer on the phase space only for $s>1/2$. As it is well known, when an integrable system is made chaotic by adding a perturbation, chaos in the system first originates in the vicinity of this separatrix layer through the so-called homoclinic tangle \cite{Reichl,linda}. As a result, there is a very thin layer of chaos around this separatrix even for very small perturbations, reminiscent of the conservative Duffing oscillator \cite{Reichl}, which is a tilted-oscillating double well~\footnote{ The LMG model with a perturbation of the form $J_{x} \cos(\omega t)$, as opposed to perturbation of the form $J_{y} \cos(\omega t)$ used in this work, would be strictly equivalent to the conservative Duffing oscillator}. 

It should also be noted that the existence of phase transitions is closely related to the presence of unstable fixed points on the classical phase space \cite{heiss_well,Zunkovic2018}. A crucial question then stands out: given that these points are also the regions of the classical phase space where chaos first develops when a nonintegrable perturbation is added to the system, how does chaos affect our ability to simulate quantum phase transitions in a noisy quantum device? We address this question in the following subsection by analyzing the effects of a nonintegrable perturbation on the bifurcation diagrams. 

\subsection{Sensitivity and robustness to perturbations in the simulation of QPTs}
The bifurcation diagrams associated with the GSQPT and the DQPT, whose bifurcation points provide us the critical-point estimates, are shown in Figs. \ref{fig:fig4}a and b in the presence of the aforementioned chaotic perturbation for $J=100$ $(N=200)$. The black curves in these figures represent the ideal bifurcation diagrams, whereas the red and the green curve represent the perturbed cases associated with the GSQPT and the DQPT correspondingly. Note that, for the case of the GSQPT, the perturbation has significantly modified the bifurcation curve for all values of $s\gtrsim s_{c}^{(gpt)}(N)$, whereas the perturbed-DQPT bifurcation diagram has been altered significantly only in a small range of $s$ values. This difference in behavior of the perturbed bifurcation diagrams can be attributed to the way chaos emerges in the system when a nonintegrable perturbation is added. Note that the initial state associated with the GSQPT protocol is always in the immediate vicinity of the unstable fixed point and therefore very close to the separatrix layer for all $s>s_{c}^{(gpt)}(N)$. As a result, the initial state in this case is always in the chaotic sea for $s>s_{c}^{(gpt)}$ as shown by the red dot on the Poincar\'{e} sections in Figs. \ref{fig:fig4}d-f. This explains the sensitivity of the GSQPT bifurcation diagram to the perturbations for $s>s_{c}^{(gpt)}$. In contrast, the initial condition associated with the DQPT is closer to the separatrix layer only around the DQPT critical point, and therefore present in the chaotic sea only for these $s$ values as shown by the green dot in the above-mentioned figures.

The different behavior of the GSQPT and the DQPT bifurcations upon addition of a perturbation can be further corroborated by analyzing the Lyapunov exponents for the associated initial conditions as a function of $s$ as shown in Fig. \ref{fig:fig4}c, where the red curve and the green curve label the Lyapunov exponents associated with the GSQPT and the DQPT initial conditions correspondingly. It can be seen here that the GSQPT initial condition has positive Lyapunov exponents for all $s>s_{c}^{(gpt)}$, whereas the DQPT initial condition has positive Lyapunov exponents only around $s_{c}^{(dpt)}$. The Lyapunov exponents here are computed using the method introduced in \cite{Habib}. Also, note that the depth of the effective potential well increases as $s$ is increased, which means that the perturbation does not affect the potential well as much for larger values of $s$ compared to smaller $s$ values. This can also be observed in the Poincar\'{e} sections shown in Fig. \ref{fig:fig4}f, where the potential wells are still intact surrounded by a chaotic sea. This explains the robustness of the DQPT bifurcation diagram for larger values of $s$.

\begin{figure}[t] 
\includegraphics[width=\linewidth, height=7cm]{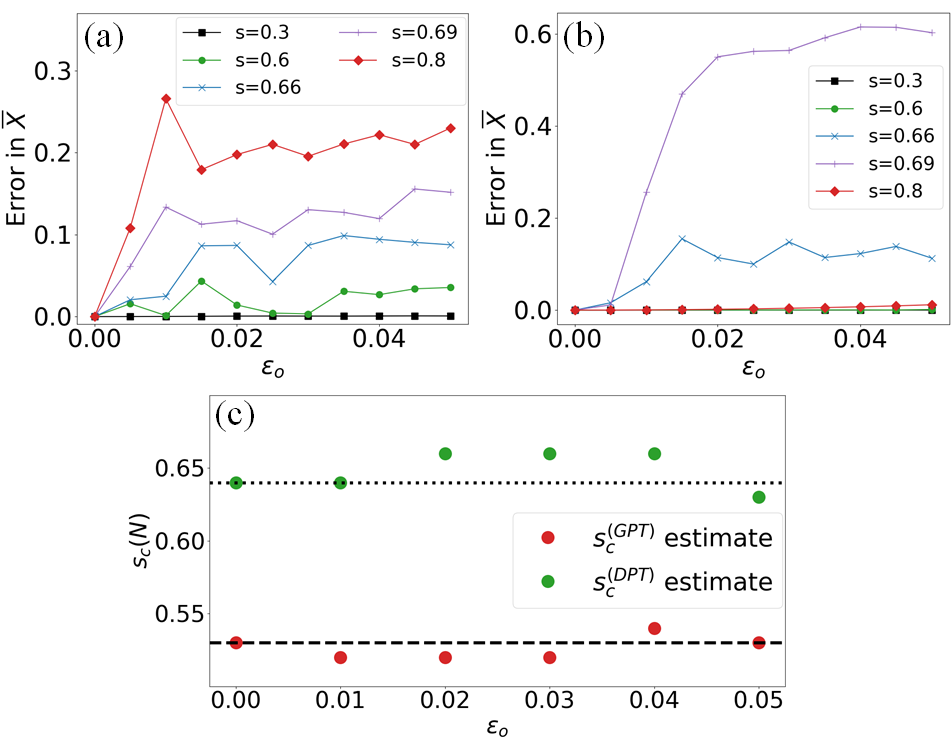} 
\caption{Time-averaged magnetization as a function of perturbation amplitude, $\varepsilon_{0}$ for different values of the control parameter, $s$, in the GSQPT(a) and the DQPT(b) protocols. (c) The critical point estimates associated with the GSQPT (green) and the DQPT (red) as a function of perturbation amplitude, $\varepsilon_{0}$. It can be seen that these estimates are robust to perturbation.}
\label{fig:fig5}
\end{figure}

So far, we've focused our analysis on a specific perturbation amplitude $\varepsilon_{0}=0.05$, but the behavior observed here is fairly general. That is, for any perturbation amplitude, there is a significant error in the time-averaged magnetization for all $s>s_{c}^{(gpt)}$ in the case of GSQPT as shown in Fig. \ref{fig:fig5}a, whereas there is a significant error in time-averaged magnetization only for intermediate values of $s_{c}^{(gpt)} < s \lesssim 0.8$ for the case of DQPT as shown in Fig. \ref{fig:fig5}b. 

The results shown in Figs. \ref{fig:fig5}a,b demonstrate that the onset of chaos makes the system more sensitive to perturbations around the critical regions. However, a crucial question is, what is the particular quantity one wants to extract from a quantum simulation?  For some applications, one is not interested in the exact value of the time-averaged magnetization, but in extracting the value of the critical point, i.e, the value of the control parameter $s$ which separates the two phases involved in the QPT. In our protocol, this can be done by identifying, for each bifurcation diagram, the value of $s$ at the time-averaged magnetization becomes larger than a small threshold $\overline{X}_{th}<<1$. The resulting critical-point estimates associated with both the quantum phase transitions are then shown in Fig. \ref{fig:fig5}c as a function of the perturbation strength, $\varepsilon_0$. Importantly, we observe that these critical point estimates are {\em not} significantly affected by the perturbation, and remain close to the ideal values even when the perturbed system is highly chaotic and the associated time-averaged magnetization have been modified significantly with respect to the ideal case. This can be understood to be a consequence of the fact that the perturbation renders the dynamics of the system chaotic only for $s\gtrsim s_c$ (for both types of QPTs). As a result, the time-averaged magnetization undergoes a drastic change, with or without the presence of the perturbation, between being zero before the transition (with dynamics largely unaffected by perturbation) to being nonzero after the transition, either due to the change in phase for $\varepsilon_o=0$ or the system turning chaotic for $\varepsilon_o>0$. In this way, the robustness of the critical point estimates can be attributed to the fact that they signal a {\em global} change in the structure of the LMG phase space, which occurs with or without the perturbation.

\begin{figure}[t] 
\includegraphics[width=1\linewidth]{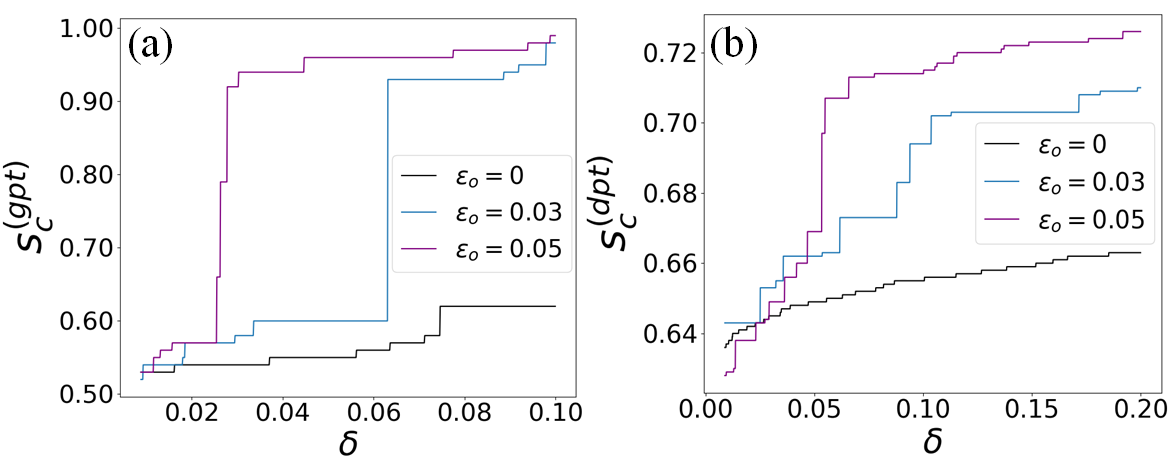}
\caption{Critical point estimates as a function of the value of the threshold parameter $\delta$, where $\overline{X}_{th}=\sin\left(\delta/\sqrt{2J}\right)$, for the GSQPT (a) and the DQPT (b). It can be seen in both cases, changing the threshold  modifies the critical-point estimate significantly.}
\label{fig: fig6}
\end{figure}

Although the perturbation changes the value of the critical point estimates slightly for both the phase transitions, it does significantly affect the sensitivity of critical point estimates to the value of the threshold $\overline{X}_{th}=\sin(\frac{\delta}{\sqrt{2J}})$ used in identifying these points. The critical point estimates as a function of threshold value are plotted in Figs. \ref{fig: fig6}a,b, where the ideal-critical-point estimates have been labelled by the black curve and the perturbed-critical-point estimates by the blue and the magenta colored curves for both the GSQPT and the DQPT. Notice that the perturbed-critical-point estimates vary over a wide range of $s$ values, particularly for the GSQPT, whereas the ideal estimates are reasonably bounded. This sensitivity could play an important role in experiments where the lack of resolution in the time-averaged magnetization might make the critical point estimates more sensitive to the perturbation.
\section{Summary and Discussion}
In this work, we have studied the effects of chaos arising from a perturbation on the quantum simulation of QPTs in the LMG Hamiltonian. One expects chaos to play an important role in the reliability of such simulations given the connection between QPTs, bifurcations, unstable fixed points, and instabilities that arise in the semiclassical behavior of an order parameter in the thermodynamic limit.  We have proposed a protocol that allows us to extract the critical points of both the GSQPT and the DQPT by measuring the time-averaged expectation values of the collective spin operators. This is motivated by the dynamics in the semiclassical limit, which can be described by the motion of a fictitious particle in a one dimensional potential well. Both the GSQPT and the DQPT are intuitively understood in terms of the motion in the potential well.
Given the integrability in the unperturbed case, we are able to derive an analytic expression for the time-averaged magnetization, which describes the output  of the ideal simulation in the thermodynamic limit. This protocol also allows us to estimate the finite-size critical point in a finite-dimensional system provided the appropriate time-averaging is performed.

To study the effect of chaos, we added a weak external time-dependent perturbation during the simulation of the LMG Hamiltonian, which becomes chaotic in a manner reminiscent of the conservative Duffing oscillator. The time-averaged magnetization associated with both the GSQPT and the DQPT are affected significantly, particularly at the values of the control parameter $s$ closer to the critical point. Furthermore, we have shown that the range of $s$ values where time-averaged magnetization have been affected the most can be identified based on the location of the initial state of the protocol with respect to the separatrix layer on the classical phase space, as the homoclinc tangle forms in the transition to chaos. Finally, despite of the sensitivity of the time-averaged magnetization to the presence of the chaotic perturbation, we showed that the critical point estimates, obtained from the analysis of the perturbed evolution, are robust for both the GSQPT and the DQPT. This robustness can be attributed to the fact that the critical points signal a change in the global structure of the LMG phase space that is also captured by the perturbed evolution.

We have studied here the effect of chaos on quantum simulation of GSQPT and DQPT for the simplest case, the LMG model, a completely connected Ising model with two-body interactions. In future, it would be interesting to explore the effect of chaos on the signatures of ESQPT. Given that the states undergoing ESQPTs are localized around the unstable fixed point, we would expect the associated signatures of these phase transitions to be fragile to the presence of the chaos, particularly in the vicinity of their critical points. In addition, we also plan to analyze more complex many-body systems.  A first rich generalization is to analyze $p$-spin models, with  all-to-all $p$-body interactions \cite{Jorg2010,Bapst2012,Munoz2021}. In contrast to the case $p=2$ (the LMG model), the cases of $p>2$ exhibit discontinuous first-order QPTs and are characterized by a gap that closes exponentially with $N$~\cite{Jorg2010}, making quantum simulation more challenging in some protocols.  Moreover, it has been shown that the phase-space associated with the kicked-version of these models undergo rapid changes at various bifurcation points compared to the LMG model \cite{Munoz2021}. We expect these new classes of dynamical instabilities to impact the reliability of the quantum simulation of $p$-spin models. Beyond quantum mean-field models, describable by a collective spin with one degree of freedom, it is important to study the impact of many-body chaos on more general many-body models. 

In the context of NISQ quantum simulators that do not have access to error correction, it is important to identify quantities that are robust with respect to the error but also hard to simulate classically~\cite{Deutsch2012, Preskill, Deutsch2020}. In this context, what is the relationship between the hardness of quantum simulation and its robustness to perturbations \cite{Deutsch2012, Deutsch2020}?  In particular, are computationally hard analog quantum simulations limited by many-body chaos, while computationally simulable problems are robust to dynamical perturbations?  The work presented here gives some initial indications in this direction. Identifying the quantum critical points is robust in this case because these correspond to changes in global structure of phase space, which was possible to identify even without keeping track of the exact quantum state. On the other hand, other quantities that depend on simulating the exact trajectory are fragile.  In a general many-body system, the question is then, which structural changes in the effective phase space can be efficiently simulated, and which emergent critical phenomena cannot, and how robust are they in the presence of dynamical instabilities?

\section{Acknowledgements}
We thank Poul Jessen and Manuel Mu\~{n}oz for numerous insightful discussions. This work was supported by the U.S. National Science Foundation under grant numbers PHY-1820679 and PHY-1630114.

\appendix
\section{Finite-size scaling} \label{sec:FSS}
In this subsection, we derive an analytic expression for the scaling of the critical point with the system size for the GSQPT, $s_{c}^{(gpt)}(N)$. We obtain this expression using the effective Schrodinger equation introduced in section \ref{sec:semiclassical}, which is a good approximation in the semi-classical limit. The main idea in this derivation is to identify the critical point of the GSQPT with the value of $s$ at which the zero-point energy of the double well equals the barrier height of the double-well. The potential of the semi-classical well is given by 
\begin{multline}\label{potential}
V(s,z) = -\frac {1 - s}{2} \sqrt{1 - z^2} - \frac{s}{4} z^{2} \\
 - \frac{1 - s}{4} \biggl(\frac{2}{N\sqrt{1 - z^2}} - \frac{1}{N^{2}}\frac{1 + z^{2}} {(1 - z^2)^{3/2}}\biggr)
\end{multline}
The above expression for the potential allows us to compute the barrier height as shown below 
\begin{multline}
V(s,0)-V\bigl(s,z_{min}\bigr)=\biggl(s-1+\frac{1}{4s}\biggr)+\frac{1}{N}\biggl(s-\frac{1}{2}\biggr)
\\
+\mathcal{O}\biggl(\frac{1}{N^{2}}\biggr)
\end{multline}
In addition, the zero-point energy can also be computed using the potential energy expression in Eq. (\ref{potential}). The main idea is to notice that the potential energy is of the form $V(z)=V_{0}+\frac{1}{2}m\omega^{2}z^{2}$, and therefore the zero-point energy is given by $\frac{1}{N}\frac{\omega}{2}$ with $\omega=\sqrt{\frac{V''(z_{min})}{m}}$. As a result, the zero-point energy is given by 
\begin{align}
E_{0}=\frac{h}{\sqrt{2}\sqrt{m}}\sqrt{\frac{s(2s-1)}{(s-1)^{2}}}+\mathcal{O}\biggl(\frac{1}{N^{2}}\biggr)
\end{align}
Finally, setting the barrier equal to the zero-point energy, one obtains
\begin{align}
s_{c}^{(gpt)}(N) &= \frac{1}{2} + \frac{N^{-\frac{2}{3}}}{2^{\frac{1}{3}}} +\mathcal{O}\biggl(\frac{1}{N}\biggr)
\end{align}
The above equation can be rewritten as
\begin{equation}
\frac{s_{c}^{(gpt)}(N)-s_{c}^{(gpt)}}{s_{c}^{(gpt)}} \propto N^{-\frac{2}{3}}
\end{equation}
Note that the above expression agrees with the predictions of the finite-size scaling ansatz \cite{Monte}, $\frac{s_{c}(N)-s_{c}}{s_{c}}\propto N^{-\frac{1}{\nu^{*}}}$ where $s_{c}(N)$ is the effective critical point for finite-size system $N$ and $\nu^{*}$ is the exponent associated with the divergence of the coherence number at the critical point \cite{Botet}. In the case of fully connected models, this exponent $\nu^{*}$ is related to the mean-field exponent $\nu_{MF}$ through the upper critical dimensionality of the corresponding finite-ranged model, $\nu^{*}=d_{c}v_{MF}$.
For the case of LMG, $d_{c}=3$ and $v_{MF}=\frac{1}{2}$, therefore $v^{*}=\frac{3}{2}$ \cite{Botet}. Note that this result $v^{*}=\frac{3}{2}$ is also consistent with the results in \cite{Botet, OTOC,joseph2_well,susceptibility}.

\section{Classical bifurcation} \label{sec:classical bifu}
\begin{figure}[h!]
\includegraphics[width=2.5in]{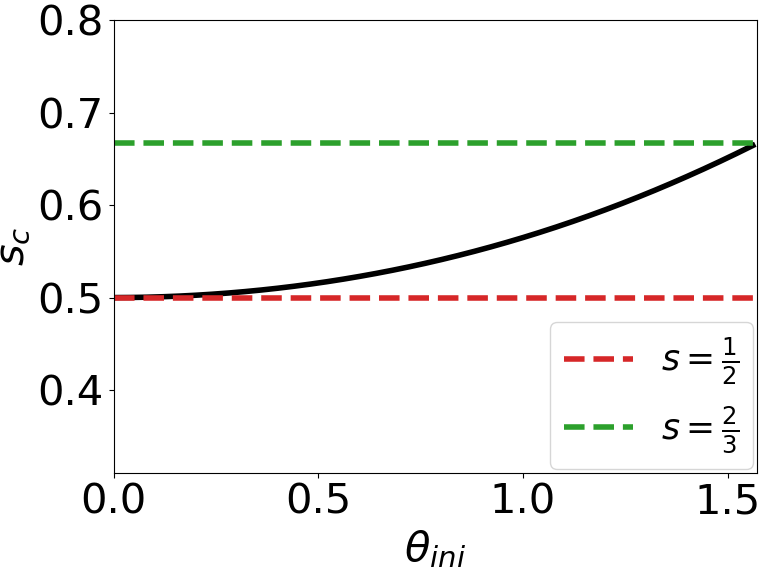}
\caption{ Bifurcation point as a function of $\theta_{ini}$, c.f. Eq. (\ref{ec:bifu_point}). An initial condition with $\theta_{ini}$ close to zero bifurcates at $s=0.5$ while the initial condition corresponding to $\theta_{ini}=90^{o}$ bifurcates at $s=\frac{2}{3}$}
\label{fig:fig_app1}
\end{figure}

In this section, we first derive the critical point of the bifurcation diagrams and then derive an analytic expression for the time-averaged magnetization. Given an initial condition of the form $(\theta=\theta_{0},\phi_{0}=0)$, energy conservation leads to the following relationship between $\theta(t)$ and $\phi(t)$:
\begin{equation}
\cos^2{\phi(t)}=\frac{2(1-s)(\cos(\theta_{0})-\cos(\theta))+s \sin^2{\theta_{0}}}{s \sin^2{\theta(t)}}\\
\end{equation}
The solutions to the above equation when $\phi(T)=0$  are given by \begin{equation}
\cos{\theta(T)}=\cos{\theta_{0}}
\quad \text{and}\quad
\cos{\theta(T)}=\frac{2(1-s)}{s}-\cos{\theta_{0}}
\end{equation}
The second solution should exist only after the formation of the double well on the phase space. Therefore, the critical point of the bifurcation associated with the time-averaged magnetization should be
\begin{equation}
    s_{c}=\frac{1}{1+\cos^{2}{\frac{\theta_{0}}{2}}}
    \label{ec:bifu_point}
\end{equation}
which is the value of $s$ when the second solutions starts to exist. It is interesting to note that the bifurcation point corresponding to GSQPT is more robust to deviations in $\theta_{ini}$ than the one for DQPT as shown in Fig. \ref{fig:fig_app1}.

For a given initial condition, the analytic expression for the time-averaged magnetization, $\overline{X}$, can be obtained by averaging $X(t)$ over one time period of the classical trajectory as shown below
\begin{equation}
\\\overline{X}	=\frac{1}{T}\int_{0}^{T}dtX(t)
	=\frac{1}{T}\int_{0}^{T}dt\sin\bigl(\theta(t)\bigr)\cos\bigl(\phi(t)\bigr)
\end{equation} 
where $\theta(t)$ and $\phi(t)$ are the spherical coordinates of the associated trajectory. The classical equations of motion for the LMG Hamiltonian in the spherical coordinates are given by 
\begin{align}
\frac{d\theta}{dt}	&=\frac{s}{2}\sin(\theta)\sin(2\phi) \\
\frac{d\phi}{dt}	&=-(1-s)+s\cos(\theta)\cos^{2}(\phi)
\end{align}
Using the above equations of motion, the integrand can be expressed as $dt=\frac{d\theta}{\frac{s}{2}\sin(\theta)\sin(2\phi)}$. Also conservation of energy allows all the $\phi(t)$ terms in the integral to be expressed as a function of $\theta(t)$. Evaluating the integral for an initial condition of the form $(\theta=\theta_{0},\phi=0)$ we obtain,
\begin{equation}
\overline{X}=- \frac{\pi}{sT}
\end{equation}
The time period, $T$, can then be evaluated as follows
\begin{align}
T&=	\int_{0}^{T}dt=\frac{1}{s}\int_{\theta_{0}}^{\theta(T)}\frac{d\theta}{\sin(\theta)\cos(\phi)\sin(\phi)} \\
\begin{split}
T&=	\frac{1}{s}\frac{2i}{\sin\theta_{0}}\biggl[F\biggl(\frac{\pi}{2}\biggl\vert1+\frac{2a}{sin^{2}\theta_{0}}\Delta z\biggr) 
	\\
    &-F\biggl(\sin^{-1}\bigl(\frac{1}{\sqrt{1+\frac{2a}{sin^{2}\theta_{0}}\Delta z}}\bigr)\biggl\vert1+\frac{2a}{sin^{2}\theta_{0}}\Delta z\biggr)\biggr]
\end{split}
\end{align}
Therefore,
\begin{equation}
\overline{X}=\mp\frac{\pi}{2}\frac{\sin\theta_{0}}{i\biggl[F\biggl(\frac{\pi}{2}\biggl\vert\lambda(\theta_{0},s)\biggr)-F\biggl(\\sin^{-1}\bigl(\frac{1}{\sqrt{\lambda(\theta_{0},s)}}\bigr)\biggl\vert\lambda(\theta_{0},s)\biggr)\biggr]}
\end{equation}
where $\lambda=1+\frac{4(1-s)}{s\sin^{2}\theta_{0}}\bigl(\cos(\theta_{0})-\frac{1-s}{s}\bigr)$. The above formula is valid only for $s>\frac{1}{1+\cos^{2}\frac{\theta_{0}}{2}}$ because the limits of the integral were chosen assuming this condition holds true. Note that $\mp$ has been added to the above expression to account for the fact that the direction of trajectory switches as the fixed point passes through the initial condition, which happens at $s=\frac{1}{1+\cos\theta_{0}}$. Using various identities, the above expression can be reexpressed as 
\begin{equation}
\overline{X}=\frac{\pi}{2} \frac{\sin{\theta_{0}}}{K\bigl(\Lambda(\theta_{0},s)\bigr)}
\end{equation}
where $K$ denotes the complete elliptic integral of the first kind and $\Lambda(\theta_{0},s)=-\frac{4(1-s)}{s\sin^{2}(\theta_{0})}\bigl(\cos(\theta_{0})-\frac{1-s}{s}\bigr)$

\end{document}